# Enhancement of superconducting properties in FeSe wires using a quenching technique

Toshinori Ozaki[1,4], Yoshikazu Mizuguchi[2,4], Satoshi Demura[1,3,4], Keita Deguchi[1,3,4], Yasuna Kawasaki[1,3,4], Toru Watanabe[1,3], Hiroyuki Okazaki[1,4], Hiroshi Hara[1,4], Hiroyuki Takeya[1,4], Takahide Yamaguchi[1,4], Hiroaki Kumakura[1,3,4], and Yoshihiko Takano[1,3,4]

[1] National Institute for Materials Science, 1-2-1 Sengen, Tsukuba, Ibaraki 305-0047, Japan

[2] Tokyo Metropolitan University, 1-1 Minami-osawa, Hachioji 192-0397, Japan

[3] University of Tsukuba, 1-1-1Tennnodai, Tsukuba, Ibaraki 305-0047, Japan

[4] JST, Transformative Research-project on Iron Pnictides, 1-2-1 Sengen, Tsukuba, Ibaraki 305-0047, Japan

(Abstract)

Enhancements of superconducting properties were observed in FeSe wires using a quenching technique. Zero resistivity was achieved at about 10 K in quenched wires, which is about 2 K higher than that of polycrystalline FeSe bulk. Furthermore, transport $J_c$ of quenched wires showed three times higher than that of furnace-cooled wires. In contrast, the quenched polycrystalline FeSe bulks did not show the enhancement of $T_c$. The quenching technique is a greatly promising for fabricating FeSe wires with high $T_c$ and high $J_c$, and quenched FeSe wires have high potential for superconducting wire applications.

## I. Introduction

Since the discovery of superconductivity in an iron-based superconducting materials family[1-5], great efforts have been devoted to the preparation of films[6-8] and wires[9-15] for practical applications. Among these iron-based superconductors, FeSe family such as FeSe, $FeTe_{1-x}Se_x$ and $FeTe_{1-x}S_x$ has several practical advantages over the pnictide. FeSe family has the simplest structure and less toxicity compared to the other As-based compounds, which simplifies its synthesis and handling. They also possess a high upper critical field $H_{c2}$ and extremely low $H_{c2}$ anisotropy[16]. In $FeTe_{0.5}Se_{0.5}$ thin films, $J_c$ exhibits a superior high field performance up to 35 T over those of low temperature superconductors[8]. All of these aspects give reasons for practical interest as well as intrinsic scientific interest. Furthermore, FeSe with $T_c^{zero}$ ~8 K and $T_c^{onset}$ ~10 K shows that the application of pressure leads to a significant enhancement of $T_c$ up to 37 K under high pressure, the third-highest known critical temperature for any binary compound[17-19]. This is indicative of the idea of exploring if a similar effect can be induced by the compressive strain in the films and the wires. Indeed, there are many reports about pressure-induced enhancement of $T_c$ in $FeTe_{0.5}Se_{0.5}$ thin film[20-22]. Recently, we reported that the FeSe wires fabricated by the Fe-diffusion Power-in-tube PIT method showed $T_c^{zero}$ ~10.5 K, about ~2 K above that of bulk sample[23]. This may come from a unique synthetic condition: FeSe wires were heat-treated at 800°C for 2 hours and taken out from a furnace to quench them. FeSe wires with higher $T_c$ must be the greater advantage of wire applications. In this paper, we show that the superconducting properties of FeSe wires can be actually enhanced by the quenching technique.

## II. Experimental

FeSe wires were prepared by the Fe-diffusion (PIT) method. The details of the Fe-diffusion PIT method were described in Ref. 23 and 24. The Se powder was packed into pure Fe tube. The tube was drawn down to final diameter of 1.1 mm. The as-drawn wires were cut in ~5 cm long pieces, and then sealed into a quartz tube evacuated and backfilled with Ar gas. The sealed FeSe wires were heat-treated at 800°C for 2 hours. Figure 1 shows heat-treatment conditions of the sealed FeSe wires, including four kinds of heating and cooling process: slow-heating and furnace-cooling, rapid-heating and furnace-cooling, slow-heating and quenching, and rapid-heating and quenching (named

W1, W2, W3 and W4, respectively).

Polycrystalline bulk samples with a nominal composition of FeSe were prepared by conventional solid-state reaction from powders of Fe (99.9%) and Se (99.999%). The mixed powders were sealed into evacuated quartz tubes, heated at 680ºC for 15 hours, and furnace-cooled to room temperature. After that, the products were ground, pressed into pellets, and sealed into the quartz tubes with Ar gas. The sealed FeSe bulks were sintered at two conditions as follows: 680ºC for 15 hours (same heat-treatment as the bulk synthesis[17]) and 800ºC for 2 hours (same heat-treatment as the wire fabrication), and followed by quenching to room temperature, respectively.

The crystal structure was characterized by X-ray diffraction (XRD) using Cu Kα radiation. Transport critical currents ($I_c$) were measured by a standard four-probe method in liquid helium and applied magnetic fields. The magnetic field was applied perpendicularly to a longitudinal direction of the wires. The $J_c$ was calculated by dividing $I_c$ by the cross sectional area of the FeSe core excluding the hole, which was measured by optical microscope. Temperature dependence of electrical resistivity ($\rho$) was measured by the four-probe method in the temperature range of 300-2 K with a physical property measurement system (PPMS; Quantum Design). Magnetic susceptibility measurements were performed using superconducting quantum interference device (SQUID) magnetometer with an applied field of 5 Oe.

## III. RESULT AND DISCUSSION

Figure 2 shows XRD patterns of the $\omega$-$2\theta$ scan for the reacted layer obtained from FeSe wires, which were heat-treated at the four kinds of heating and cooling processes. Theses patterns indicate that all of the samples were indexed using the *P*4/*nmm* space group with small amount of hexagonal FeSe and iron-oxide impurity phase. This indicates that there is not a structure-transition caused by these heat-treatment conditions. The lattice constants *a* and *c* were plotted in fig. 3(a) and 3(b). Whereas no significant trends were observed for the *a*-axis length, a strong correlation of the *c*-axis length with the heat-treatment condition existed. Interestingly, the *c*-axis lattice constants of both quenched wires are slightly smaller than that of polycrystalline FeSe bulks, while the lattice constants *c* of both furnace-cooled wires are almost the same as the polycrystalline FeSe bulks[17]. This could be concluded that the quenched wires were compressed along the $c$-axis.

Figure 4 shows the temperature dependence of the resistivity for the four kinds of FeSe wires. The resistive transition is very sharp, with a width (10% - 90%) of about 1 K. The values of $T_c^{zero}$ of all the samples are listed in the inset of fig. 4. There is a striking evidence that $T_c^{zero}$ of FeSe wires depends on whether they are quenched or furnace-cooled. The $T_c$ values of two quenched wires are ~2 K higher than those of the polycrystalline FeSe bulks, while two furnace-cooled wires exhibit almost the same $T_c$ than those of the polycrystalline samples[17,25]. This is not the first example on the improvement of $T_c$ in iron chalcogenide. There are many reports about the $T_c$ enhancement in $FeSe_{1-x}Te_x$ thin films, due to the shrinkage of lattice parameter $c$[20-22]. We have also found that the $T_c$ values of iron-based superconductors were strongly correlated with the anion height from Fe layer[26]. From the above results, it should be understood that the enhancement of $T_c$ in quenched FeSe wires is clearly related to the shrinkage of the lattice constant $c$, arising from compressive strain.

Figure 5 shows the temperature dependence of the magnetization of the FeSe powders obtained from W2 and W3 in zero-field cooling (ZFC) and field cooling (FC) under 5 Oe magnetic field, respectively. The magnetization is shifted to a positive value, because these powders contain iron-oxide. $T_c^{mag}$ ($T_c$ estimated from the magnetization) of superconducting transition were estimated 9.5 and 10.8 K for W2 and W3, respectively. Similar to the result of the temperature dependence of the resistivity, $T_c^{mag}$ of the quenched wires is higher than that of the furnace-cooled wires. It is notable that the shielding volume fraction of W3 is rather larger than that of W2, indicating that the quenching technique increases the shielding volume fraction in FeSe wires.

Figure 6 shows the magnetic field dependence of the transport $J_c$ for the four kinds of FeSe wires. The self-field $J_c$ ($J_c^{s.f.}$) values for W1, W2, W3 and W4 showed 116, 95, 383 and 350 A/cm$^2$, as listed in the inset of fig. 6. Remarkably, The $J_c^{s.f.}$ values of the quenched wires are about three times higher than that of the furnace-cooled wires. Considering the result of the temperature dependence of the resistivity and the magnetization, we can conclude that these enhancements of $J_c^{s.f.}$ values are due to the higher $T_c$ and the larger shielding volume fraction. The $J_c$ for all the wires showed a rapid decrease at low fields and the gradually decreased with increasing magnetic fields. This indicates that the FeSe wires have an advantage for practical applications in high magnetic fields.

In order to clarify the origin of the enhancement of $T_c$ for the FeSe wires, we investigate the influence of polycrystaline FeSe bulks on the quenching technique. Figure 7 shows the temperature dependence of the resistivity for polycrystalline FeSe bulks quenched at 680°C and 800°C. For comparison, the polycrystalline FeSe bulk furnace-cooled from 680°C is plotted in the same figure. The inset gives the whole $\rho$-$T$ curves of these samples from 2 to 300 K. The $T_c^{zero}$ of these samples are also listed in the inset. Surprisingly, we did not observe any sign of the enhanced $T_c$ for quenched bulk samples. The $T_c^{zero}$ of the bulk sample quenched at 680°C is almost the same value of ~8 K as the bulk sample furnace-cooled from 680°C. The $T_c^{zero}$ of the bulk sample quenched at 800°C is slightly lower than that of the other two samples. The reason why $T_c^{zero}$ of quenched FeSe wires increased has not been clarified yet. We can be sure that both the Fe sheath surrounding FeSe and the quenching technique induce enhancements of $T_c$ and $J_c$ in FeSe wires. These results suggest that the FeSe wires using the Fe-diffusion PIT method and the quenching technique could be potential for superconducting wire applications.

## IV. CONCLUSION

The fast quenching technique improved superconducting properties of the FeSe wires fabricated by the Fe-diffusion PIT method. The quenched FeSe wires showed a zero resistivity at more than 10 K, about ~2 K above those of the furnace-cooled FeSe wires and polycrystalline FeSe bulks. The structural analysis suggests that the quenched wires are compressed along *c*-axis. The $J_c$ values of the quenched wires also present three times higher than that of furnace-cooled wires. However, quenched polycrystalline FeSe bulks did not show the enhancement of $T_c$. These results indicate that the Fe-diffusion PIT method and the quenching technique are promising for fabrication of FeSe superconducting wires, which shows high $T_c$ and $J_c$ for practical applications.

"Acknowledgments"

This work was supported in part by the Japan Society for the Promotion of Science (JSPS) through Grants-in-Aid for JSPS Fellows and 'Funding program for World-Leading Innovative R&D on Science Technology (FIRST) Program'.

(references)

(Captions)

Fig. 1 Schematic illustration of the thermal profile to fabricate FeSe wires including four kinds of heating and cooling process.

Fig. 2 X-ray diffraction patterns of four kinds of FeSe wires. ♦ indicates peaks from hexagonal phase. * indicates peaks from Fe oxide.

Fig. 3 Lattice constants (a) *a* and (b) *c* for reacted layer of four kinds of FeSe wires.

Fig. 4 Temperature dependence of the resistivity of four kinds of FeSe wires. The inset displays the values of $T_c^{zero}$ of these wires.

Fig. 5 Temperature dependence of the magnetization of the powders obtained from the FeSe wires of W2 and W3 under a magnetic field of 5 Oe by ZFC and FC conditions.

Fig. 6 Magnetic field dependence of the transport $J_c$ at liquid helium temperature (4.2

K) up to 12 T for four kinds of FeSe wires fabricated by the Fe-diffusion PIT method. The magnetic field was applied perpendicular to a longitudinal direction of the wires.

Fig. 7 Temperature dependence of resistivity of polycrystalline FeSe bulks quenched at 680°C for 15 hours, 800°C for 2 hours, and furnace-cooled from 680°C for 15 hours. The inset shows the $\rho$-$T$ curves between 2 and 300 K.

Figure 1

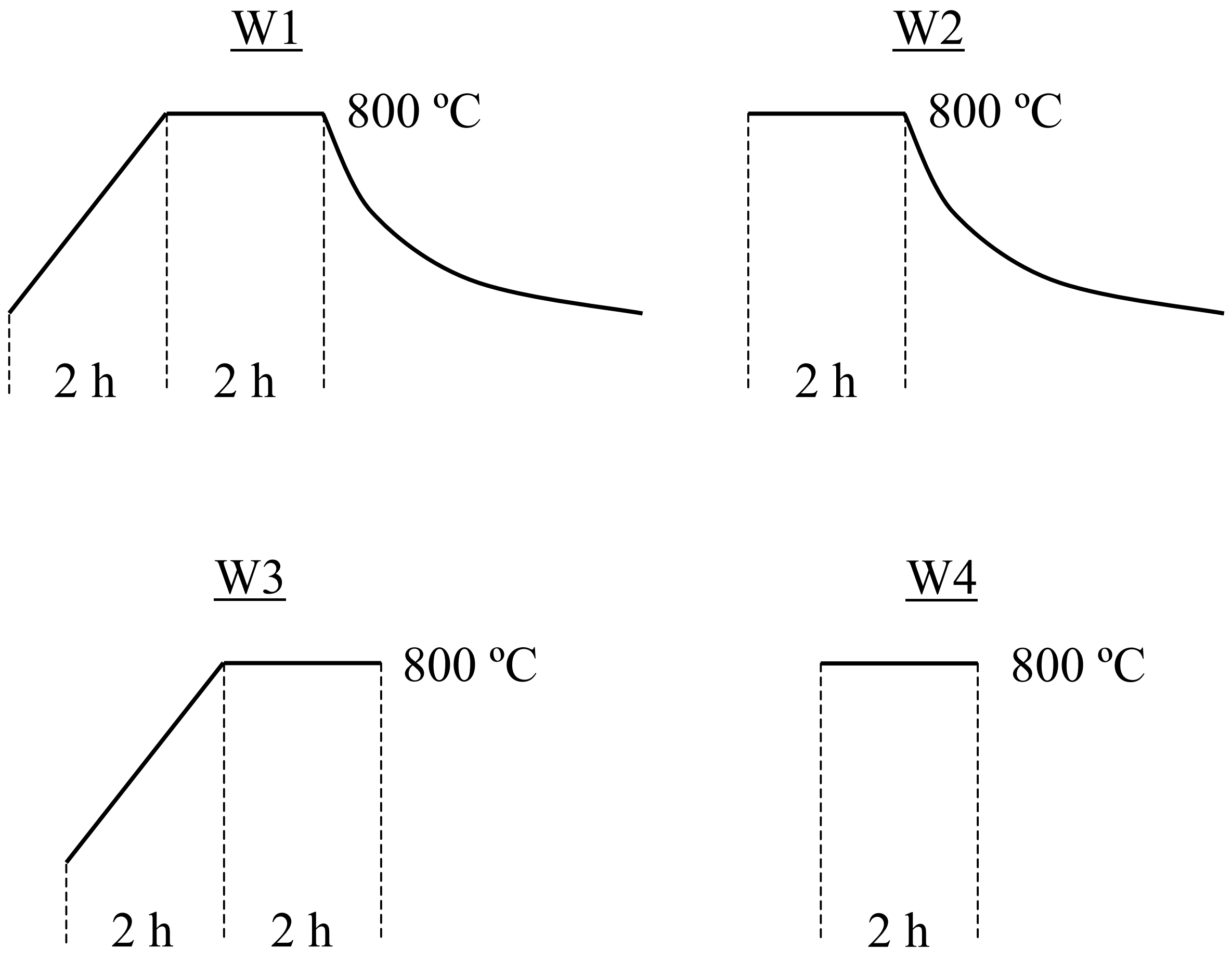

Figure 2

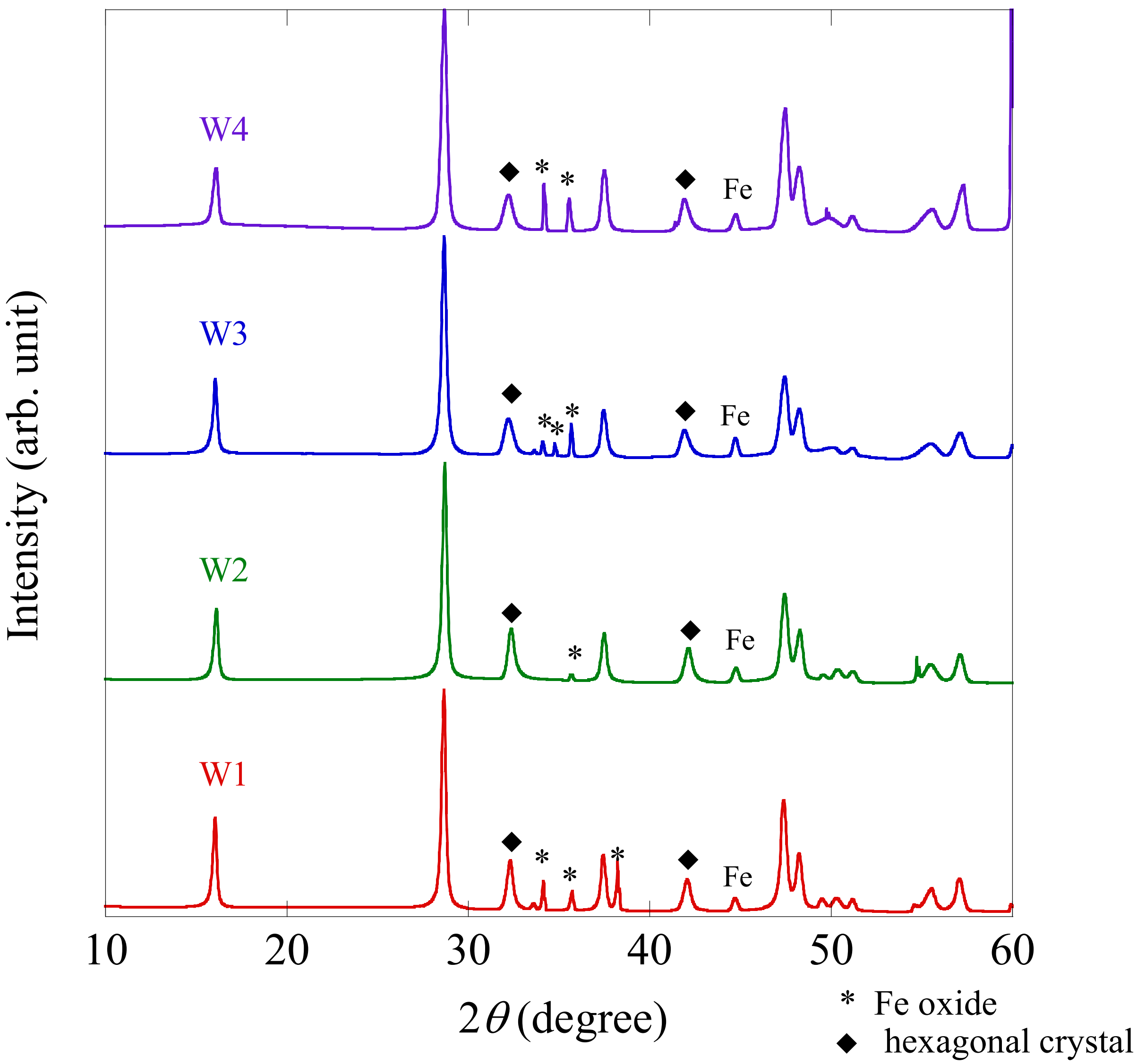

Figure 3(a)

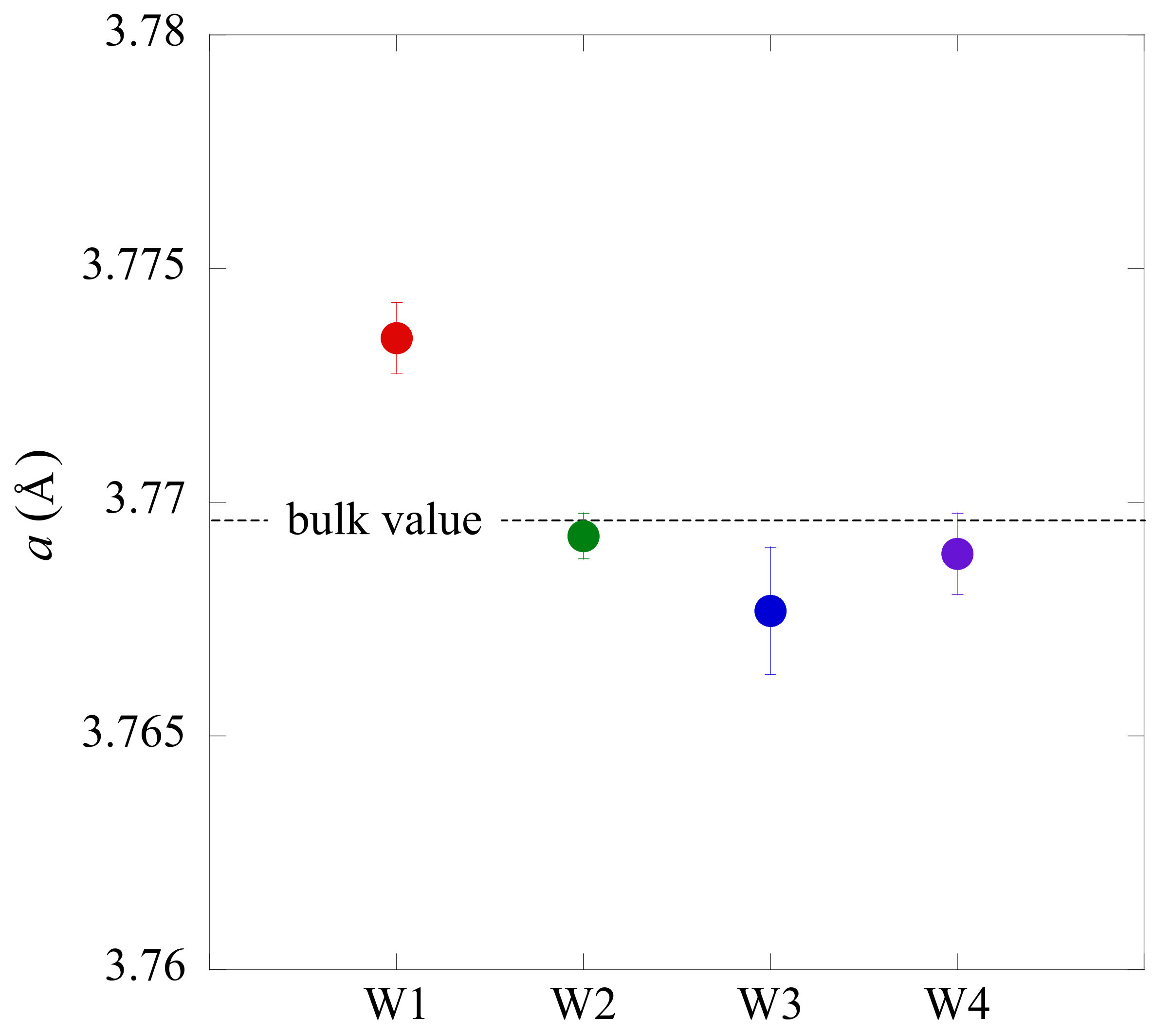

Figure 3(b)

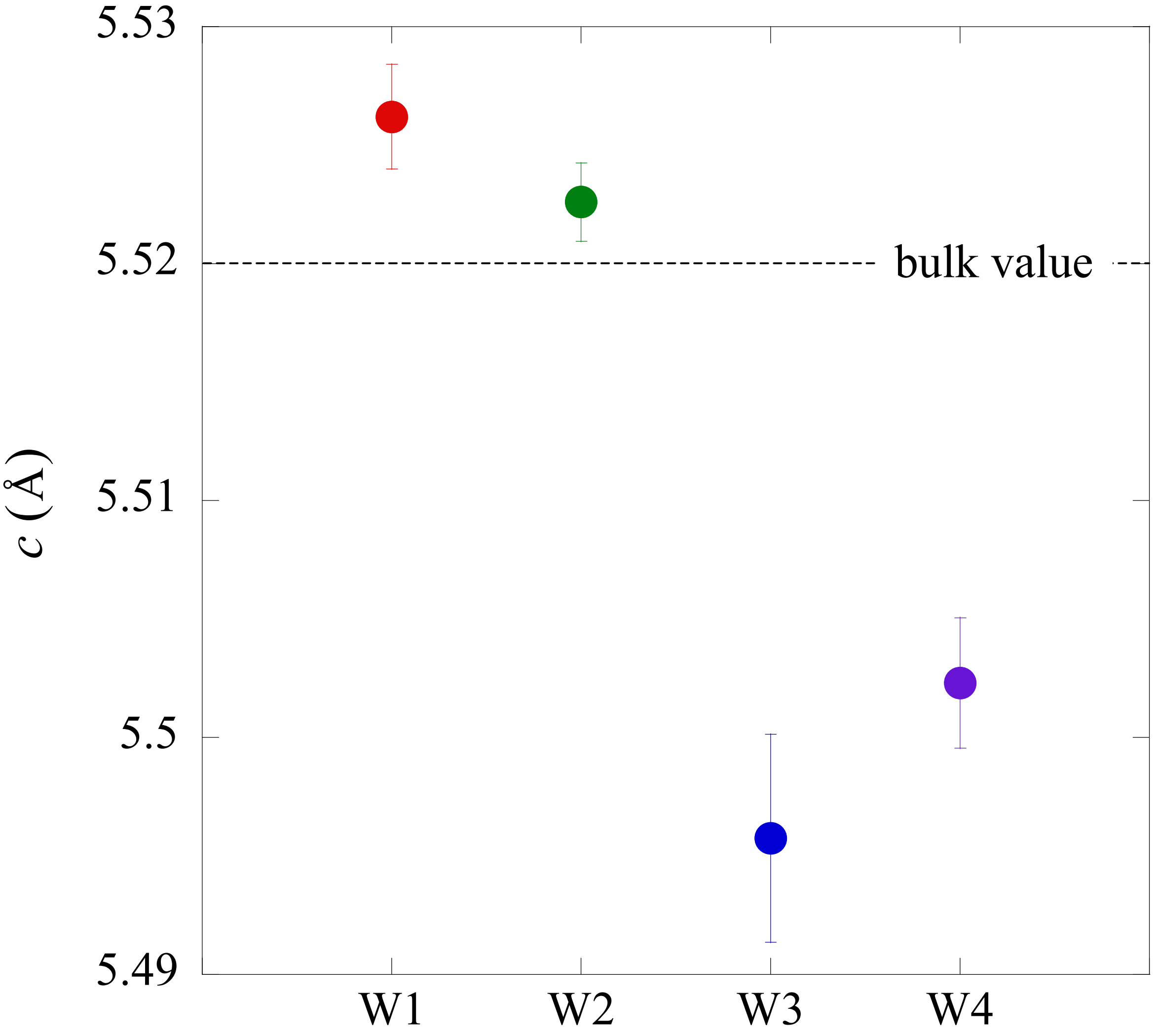

Figure 4

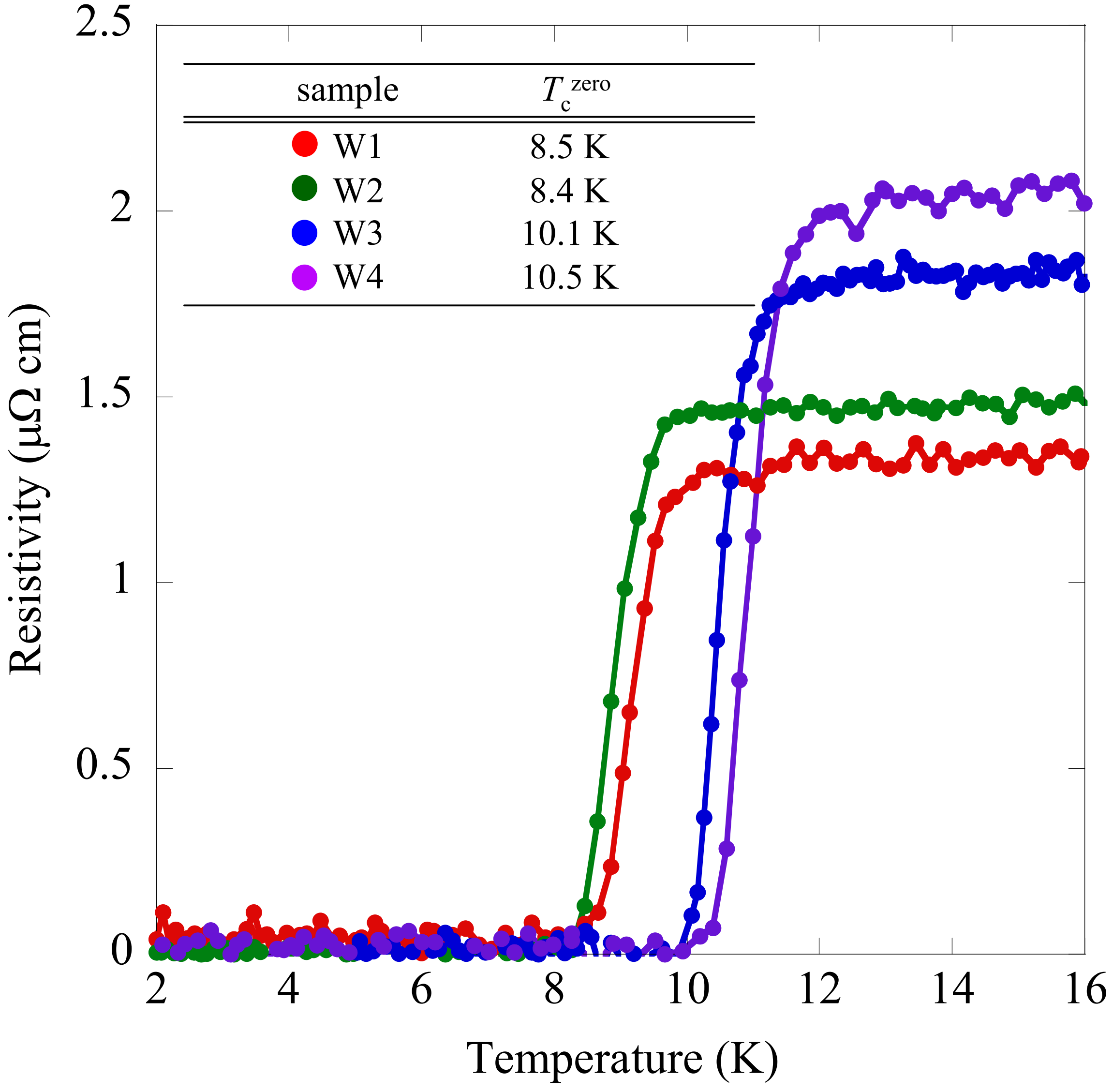

Figure 5

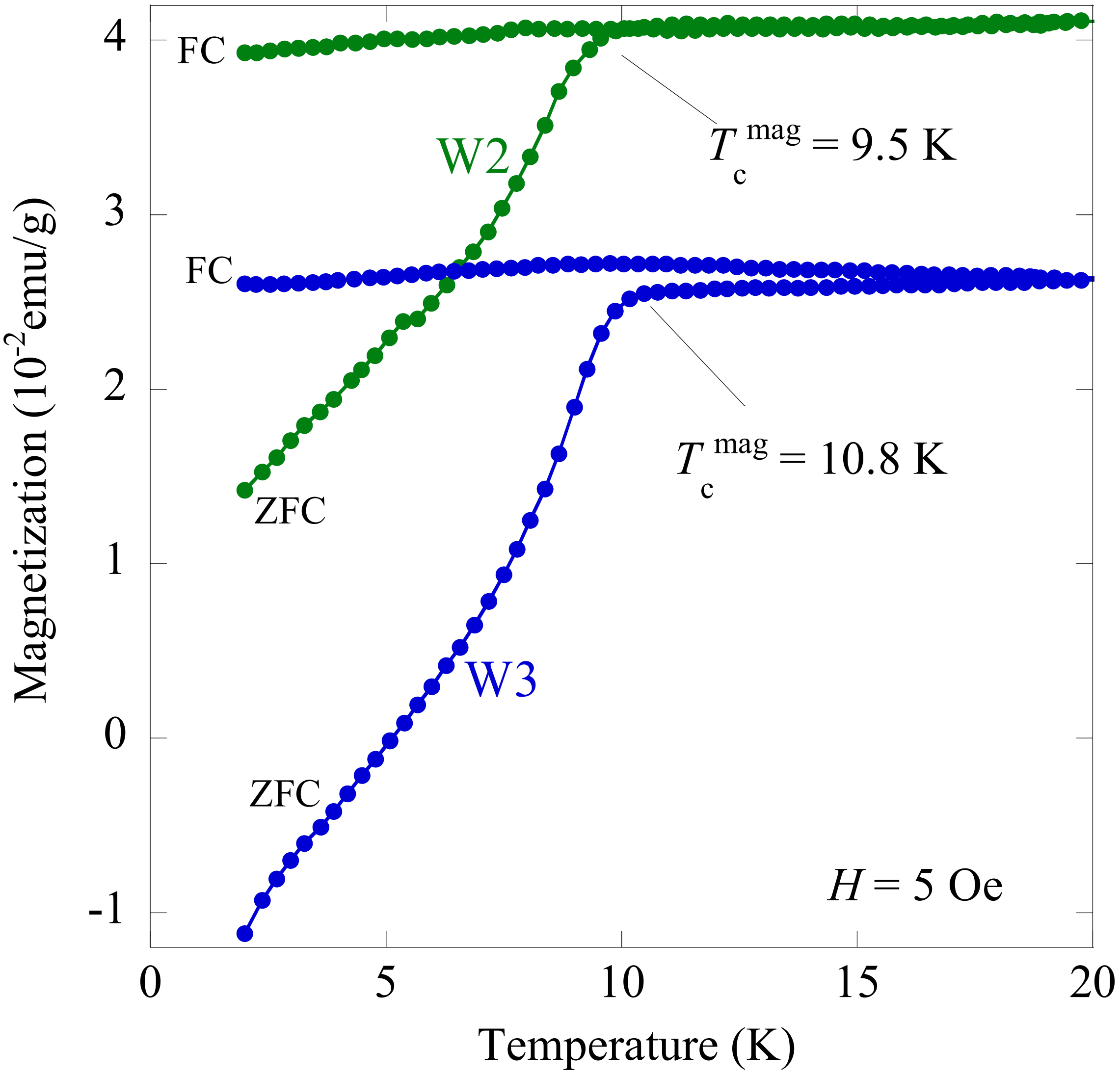

Figure 6

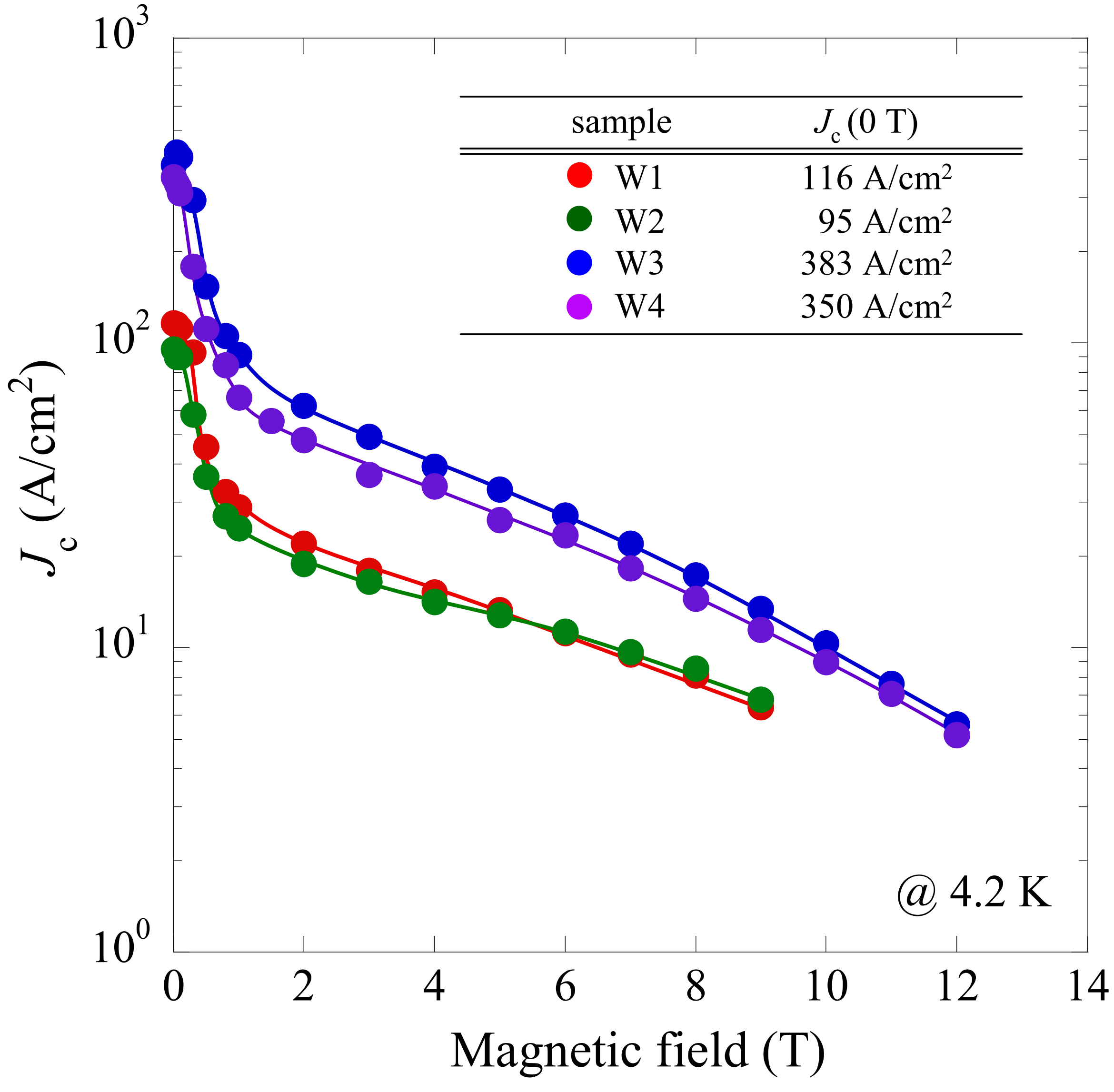

Figure 7

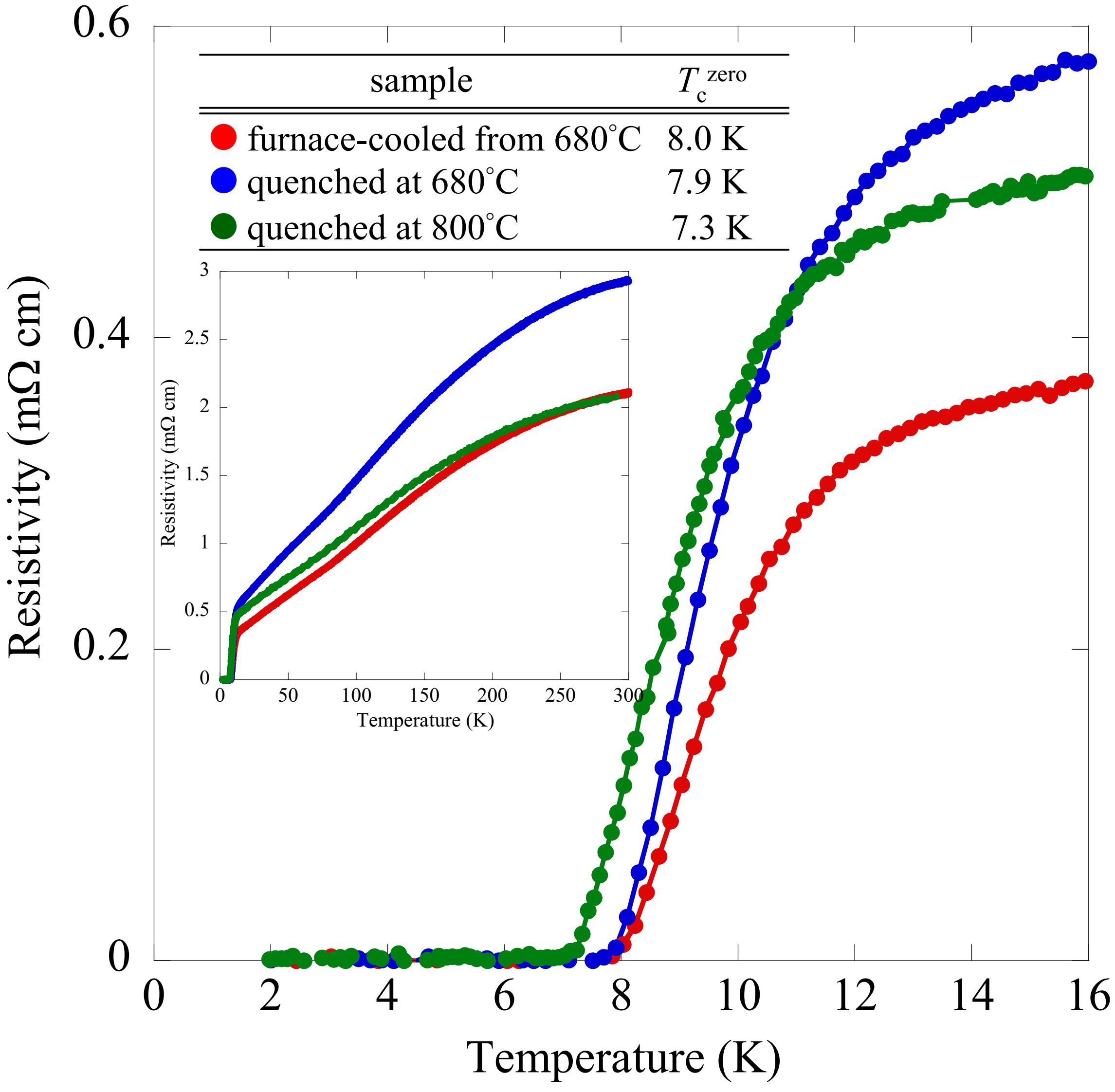